\documentclass{article}
\usepackage{graphicx}
\usepackage{amsmath}
\usepackage{subfigure}

\usepackage[all]{xy}
\usepackage{amsfonts}
\usepackage{amssymb}
\usepackage{amscd}
\usepackage{amsthm}
\usepackage{latexsym}
\usepackage{amsbsy}

\newtheorem{theorem}{Theorem}[section]
\newtheorem{remark}[theorem]{Remark}

\newtheorem{lemma}[theorem]{Lemma}

\newtheorem{definition}[theorem]{Definition}

\newcommand{\rnc}[2]{\renewcommand{#1}{#2}}
\rnc{\theequation}{\thesection.\arabic{equation}}
\rnc{\[}{\begin{equation}}
\rnc{\]}{\end{equation}}

\begin{document}

\title{Solutions to the constant Yang-Baxter equation in all dimensions}
\author{Arash Pourkia\footnote{arash.pourkia@aum.edu.kw (please cc apourkia@gmail.com)}\\Department of Mathematics, \\ American University of the Middle East, Kuwait}

%\date{\today}
\maketitle

\begin{abstract}
We will present solutions to the constant Yang-Baxter equation, in any dimension $n$. More precisely, for any $n$, we will create an infinite family of $n^2$ by $n^2$ matrices which are solutions to the constant Yang-Baxter equation. The total number of non-vanishing entries of such a matrix is $4n^2$ for $n$ even, and $4(n-1)(n)+1$ for $n$ odd. We will also present the unitary conditions for those matrices. Moreover, we discuss the entangling property of those matrices.
\end{abstract}

{\textbf{Keywords:}} Yang-Baxter equation, Quantum logic gates, Quantum entanglement, Topological entanglement.

\section{Introduction} \label{Introduction}
Yang-Baxter equation and its solutions \cite{yang 67, baxter 78, onsager 44}, play fundamental role in several areas of Physics and Mathematics \cite{perk 06, jimbo book 87} (and references therein).

 More recently, within the area of  quantum computing \cite{barenco etal 95}, on one hand the idea of topological quantum computation, and on the other hand interesting relations between quantum entanglement and topological entanglement have emerged \cite{Kitaev etal 03,freedman etal 08,aravind 97, kaufman lomonaco 02}. In particular, unitary solutions to Yang-Baxter equation  are crucially important in exploring the relations among quantum entanglement and topological entanglement \cite{kaufman lomonaco 04, kaufman etal 05-1,kaufman etal 05-2,  jchen et al 07,lho et al 10,pinto et al 13,arash-josep-2016,arash arxive 17, ZhangK ZhangY 17}. Because on one hand, (entangling) unitary solutions, are (universal) quantum logic gates, for quantum computing. On the other hand they provide representations of braid group and yield invariants of links and knots \cite{jones1985,kauffman1987,turaev 1992,kasslbook}. These aspects of Yang-Baxter equation are of particular interest to the author.  \newline

Let us recall that, for a $n$ dimensional complex vector (Hilbert) space, $V$, for which a basis has been fixed, any linear map $R:V\otimes V \rightarrow V\otimes V$ can be represented as a $n^2$ by $n^2$ matrix with complex entries (and vice versa). The  map $R$, is said to be a solution to the constant Yang-Baxter equation, if it satisfies the following relation,
\[ (R\otimes I)(I\otimes R)(R\otimes I)=(I\otimes R)(R\otimes I)(I\otimes R) \label{byb}\]
where $I$ is the identity map on $V$. Equation \eqref{byb} sometimes is called the braided Yang-Baxter equation, or simply the braid equation.\newline

It is well known that when $R$ is a solution to \eqref{byb}, composing $R$ with the swap map $S$ gives, $\hat{R}=RS$ (or $SR$),  a solution to the, so called, quantum (or algebraic) constant Yang-Baxter equation,
\[\hat{R}_{12}\hat{R}_{13}\hat{R}_{23}=\hat{R}_{23}\hat{R}_{13}\hat{R}_{12} \label{ayb}\]
where,  $\hat{R}_{12}=\hat{R} \otimes I$, $\hat{R}_{23}=I \otimes \hat{R}$, and $\hat{R}_{13}=(I \otimes S)(\hat{R} \otimes I)(I \otimes S)$. Here, the swap map $S$, also called swap gate in quantum computing, is the linear map $S:V\otimes V \to V\otimes V$, such that, $S(u\otimes v)=v\otimes u$, for all $u$ and $v$ in $V$. Also a solution $\hat{R}$ to \eqref{ayb}, results in a solution $R=\hat{R}S$ to \eqref{byb}. Therefor there is a correspondence between solutions to \eqref{byb} and solutions to \eqref{ayb}, via the swap map $S$. For $n=2$, all solutions to \eqref{ayb}, and also all unitary solutions to \eqref{byb} have been classified \cite{hietarinta, dye 03}. But for $n \ge 3$ the problem is still open. \newline

The topic of topological entanglement in a nutshell is as follows. If $R$ is invertible, it provides an infinite family of braid group representations, which in  turn, yield some invariants of links/knots. In such a process $R$ (or $R^{-1}$) is the operator for over-crossing (or under-crossing) moves \cite{jones1985,kauffman1987,turaev 1992,kasslbook}.\newline

As for the topic of quantum entanglement in quantum computing, it could be summarized as follows. When $R$  is unitary, i.e., $ R^{-1}=R^{\dagger}$ where $R^{\dagger}$ is the conjugate transpose of $R$, then $R$ could be considered as a quantum logic gate acting on $n$-qudits $|\Phi \rangle$, which are $n^2$-component vectors in $V \otimes V$. A {\it quantum state} which, abstractly, is represented by a $n$-qudit $|\Phi \rangle$ is called to be entangled if it can not be decomposed into tensor product of two other states from $V$. $R$ is said to be an entangling quantum gate if it creates entangled states by acting on un-entangled ones. In other words, $R$ is entangling if there exist an un-entangled state $|\Phi \rangle $ such that $R|\Phi \rangle $ is entangled. A quantum gate $R$ is universal if and only if it is entangling \cite{wooter-hill-1997, kauffman entng. criteria 03, brylinskis 02, alagic et al 16}.\newline

In this paper, for any dimension $n$ we will create an infinite family of $n^2$ by $n^2$ matrices which are solutions to Yang-Baxter equation (we do this for both \eqref{byb} and \eqref{ayb}). The total number of possible non-zero entries of such a matrix is $4n^2$ for $n$ even, and $4(n-1)(n)+1$ for $n$ odd. We also discuss unitarity and entangling properties of those matrices. \newline

This paper is organized as follows. In Section \eqref{Main Result}, we will use four groups of arbitrary complex numbers $a_i$'s, $b_i$'s, $x_i$'s and $y_i$'s to define a $n^2$ by $n^2$ matrix $R_{n^2}$ (or $R$ for short). Then we will prove that $R$ is a solution to Yang-Baxter equation \eqref{byb}.
In Section \eqref{Solutions to algebraic Yang-Baxter equation}, we give a full description of $\hat{R}=SR$ which is a solution to the quantum Yang-Baxter equation \eqref{ayb}.
In Section \eqref{Unitarity and Entangling conditions for $R_{n^2}$},  We will present necessary and sufficient conditions for those matrices to be unitary. Moreover, we briefly discuss the entangling property of those matrices, which implies their universality, as quantum gates. \cite{wooter-hill-1997, kauffman entng. criteria 03, brylinskis 02, alagic et al 16}. We close this paper with a short discussion and concluding remarks, followed by acknowledgement.

\section{Solutions to Yang-Baxter equation \eqref{byb} } \label{Main Result}

In this section, First, for any $n$, we define a  $n^2$ by $n^2$ matrix $R_{n^2}$, (or $R$ for short), in Definition \ref{Rn2 def}. Then in Theorem \ref{Rn2 sol yb} we prove that $R_{n^2}$ is a solution to the Yang-Baxter equation \eqref{byb}.

\begin{definition} \label{Rn2 def}
For any $n$, we define the $n^2$ by $n^2$ matrix $R_{n^2}$, as follows. Let $a_i$, $b_i$, $x_i$, $y_i$, be four sets of arbitrary complex numbers, where the range of $i$ will be clear soon. These four sets of complex numbers will occupy certain entry positions of $R$, as follows, and the rest of $R$ will be occupied by zeros.

Let integers $1 \le t,s \le m$, where $m=\frac{n}{2}$ for $n$ even and $m=\frac{n-1}{2}$ for $n$ odd. We define integers $i$, $j$, $k$, and $l$ by,
\[i=(t-1)n+s,\quad j=(s-1)n+t \label{i j}\]
\[k=(t-1)n +(n-s+1),\quad l=(n-s)n+t \label{k l}\]
Also, we define $ \tilde{i}$, $\tilde{j}$,  $\tilde{k}$, and $\tilde{l}$ by,
\[\tilde{i}=n^2+1-i=(n-t)n+(n-s+1) \label{i tld}\]
\[\tilde{j}=n^2+1-j=(n-s)n+(n-t+1) \label{j tld}\]
\[\tilde{k}=n^2+1-k=(n-t)n+s \label{k tld}\] \[\tilde{l}=n^2+1-l=(s-1)n+(n-t+1) \label{l tld}\]
Notice that $\tilde{\tilde{i}}=i,\, \tilde{\tilde{j}}=j$, and so on.

Now for each $i=(t-1)n+s$ with $1 \le t,s \le m$, where $m=\frac{n}{2}$ for $n$ even and $m=\frac{n-1}{2}$ for $n$ odd, we set $a_i$, $b_i$, $x_i$ and $y_i$ to each occupy four entry positions of $R$ as follows,
\[R_{i,j}=R_{k,l}=R_{\tilde{k},\tilde{l}}=R_{\tilde{i},\tilde{j}}=a_i  \label{a entries} \]
\[R_{i,\tilde{j}}=R_{k,\tilde{l}}=R_{\tilde{k},l}=R_{\tilde{i},j}=b_i  \label{b entries} \]
\[R_{i,\tilde{l}}=R_{k,\tilde{j}}=R_{\tilde{k},j}=R_{\tilde{i},l}=x_i  \label{x entries} \]
\[R_{i,l}=R_{k,j}=R_{\tilde{k},\tilde{j}}=R_{\tilde{i},\tilde{l}}=y_i  \label{y entries} \]

Plus, in the case of $n$ odd (with $m=\frac{n-1}{2}$), in addition to all mentioned above, we set extra $a_i$'s and $b_i$'s (but not $x_i$'s and $y_i$'s!).  For $1 \le t \le m$, we define,
\[i=(t-1)n+(\frac{n+1}{2}),\,\, j=(\frac{n-1}{2})n+t \label{i j odd}\]
and
\[\tilde{i}=n^2+1-i=(n-t)n+(\frac{n+1}{2}),\, \tilde{j}=n^2+1-j=(\frac{n-1}{2})n+(n-t+1)\label{i j tld odd}\]
and we set,
\[R_{i,j}=R_{j,i}=R_{\tilde{j},\tilde{i}}=R_{\tilde{i},\tilde{j}}=a_i  \label{a entries odd} \]
\[R_{i,\tilde{j}}=R_{j,\tilde{i}}=R_{\tilde{j},i}=R_{\tilde{i},j}=b_i  \label{b entries odd} \]

Moreover, in the case of $n$ odd, we set $R_{\frac{n^2+1}{2},\frac{n^2+1}{2}}=x$, where $x$ is another arbitrary complex number.
\end{definition}

To give an illustration, here we present $R_{n^2}$ for $n=2$, $n=3$ and $n=4$.

{\small{
$$R_4=\left(\begin{array}{rrrr}
a_{1} & x_{1} & y_{1} & b_{1} \\
y_{1} & b_{1} & a_{1} & x_{1} \\
x_{1} & a_{1} & b_{1} & y_{1} \\
b_{1} & y_{1} & x_{1} & a_{1}
\end{array}\right)
R_9=\left(\begin{array}{rrrrrrrrr}
a_{1} & 0 & x_{1} & 0 & 0 & 0 & y_{1} & 0
& b_{1} \\
0 & 0 & 0 & a_{2} & 0 & b_{2} & 0 & 0 &
0 \\
y_{1} & 0 & b_{1} & 0 & 0 & 0 & a_{1} & 0
& x_{1} \\
0 & a_{2} & 0 & 0 & 0 & 0 & 0 & b_{2} &
0 \\
0 & 0 & 0 & 0 & x & 0 & 0 & 0 & 0 \\
0 & b_{2} & 0 & 0 & 0 & 0 & 0 & a_{2} &
0 \\
x_{1} & 0 & a_{1} & 0 & 0 & 0 & b_{1} & 0
& y_{1} \\
0 & 0 & 0 & b_{2} & 0 & a_{2} & 0 & 0 &
0 \\
b_{1} & 0 & y_{1} & 0 & 0 & 0 & x_{1} & 0
& a_{1}
\end{array}\right)$$}}

{\small{
		$$R_{16}=$$}}
{\small{
$$\left(\begin{array}{rrrrrrrrrrrrrrrr}
a_{1} & 0 & 0 & x_{1} & 0 & 0 & 0 & 0 &
0 & 0 & 0 & 0 & y_{1} & 0 & 0 & b_{1} \\
0 & 0 & 0 & 0 & a_{2} & 0 & 0 & x_{2} &
y_{2} & 0 & 0 & b_{2} & 0 & 0 & 0 & 0 \\
0 & 0 & 0 & 0 & y_{2} & 0 & 0 & b_{2} &
a_{2} & 0 & 0 & x_{2} & 0 & 0 & 0 & 0 \\
y_{1} & 0 & 0 & b_{1} & 0 & 0 & 0 & 0 &
0 & 0 & 0 & 0 & a_{1} & 0 & 0 & x_{1} \\
0 & a_{5} & x_{5} & 0 & 0 & 0 & 0 & 0 &
0 & 0 & 0 & 0 & 0 & y_{5} & b_{5} & 0 \\
0 & 0 & 0 & 0 & 0 & a_{6} & x_{6} & 0 &
0 & y_{6} & b_{6} & 0 & 0 & 0 & 0 & 0 \\
0 & 0 & 0 & 0 & 0 & y_{6} & b_{6} & 0 &
0 & a_{6} & x_{6} & 0 & 0 & 0 & 0 & 0 \\
0 & y_{5} & b_{5} & 0 & 0 & 0 & 0 & 0 &
0 & 0 & 0 & 0 & 0 & a_{5} & x_{5} & 0 \\
0 & x_{5} & a_{5} & 0 & 0 & 0 & 0 & 0 &
0 & 0 & 0 & 0 & 0 & b_{5} & y_{5} & 0 \\
0 & 0 & 0 & 0 & 0 & x_{6} & a_{6} & 0 &
0 & b_{6} & y_{6} & 0 & 0 & 0 & 0 & 0 \\
0 & 0 & 0 & 0 & 0 & b_{6} & y_{6} & 0 &
0 & x_{6} & a_{6} & 0 & 0 & 0 & 0 & 0 \\
0 & b_{5} & y_{5} & 0 & 0 & 0 & 0 & 0 &
0 & 0 & 0 & 0 & 0 & x_{5} & a_{5} & 0 \\
x_{1} & 0 & 0 & a_{1} & 0 & 0 & 0 & 0 &
0 & 0 & 0 & 0 & b_{1} & 0 & 0 & y_{1} \\
0 & 0 & 0 & 0 & x_{2} & 0 & 0 & a_{2} &
b_{2} & 0 & 0 & y_{2} & 0 & 0 & 0 & 0 \\
0 & 0 & 0 & 0 & b_{2} & 0 & 0 & y_{2} &
x_{2} & 0 & 0 & a_{2} & 0 & 0 & 0 & 0 \\
b_{1} & 0 & 0 & y_{1} & 0 & 0 & 0 & 0 &
0 & 0 & 0 & 0 & x_{1} & 0 & 0 & a_{1}
\end{array}\right)$$ }}

\begin{remark}
{\emph{Notice that, each of  $a_i$'s, $b_i$'s, $x_i$'s, first appears in row {$\textit{\textbf{i}}$} of $R$, and then re-appears in other rows as dictated by Formulas \eqref{i j}-\eqref{b entries odd}. The range of $i$, for $n$ even is $1 \le i \le (m-1)n+m$, and for $n$ odd is $1 \le i \le (m-1)n+(\frac{n+1}{2})$.		
Also, notice that, all four sets of complex numbers $a_i$'s, $b_i$'s, $x_i$'s and $y_i$'s, if taking into account their repetitions, occupy $4n^2$ entry positions of $R$  when $n$ even, and $4(n-1)(n)$ entries when $n$ odd. This means, the total number of possible non-zero entries of $R_{n^2}$ is $4n^2$ for $n$ even, and $4(n-1)(n)+1$ for $n$ odd. The latter is because of the extra entry {$\textit{\textbf{x}}$} at the center of $R$ when $n$ is odd. }}
\end{remark}

Now we prepare for an equivalent definition of $R_{n^2}$, which is less expressive than the above one, but will be very useful in the proof of our main theorem, Theorem \ref{Rn2 sol yb} below. In Definition \ref{Rn2 def}, let us see, for example, how each $a_i$ occupy four entry positions of $R$. By looking at Formulas \eqref{i j}-\eqref{l tld} and \eqref{a entries} we notice that whenever two indices from $i$, $j$,$\cdots$,$\tilde{l}$ pair together to set an entry position for any $a_i$, namely in $R_{i,j}=R_{k,l}=R_{\tilde{k},\tilde{l}}=R_{\tilde{i},\tilde{j}}=a_i$, they essentially fit the following pattern,
\[R_{v, w}=a_i \text{, where, } v=(t-1)n+s,\quad w=(s-1)n+t \label{v w} \]
with the following transformations allowed.
\[t \rightarrow n-t+1, \quad s \rightarrow n-s+1  \label{t s transf} \]
and with $1 \le t,s \le n$ (instead of $1 \le t,s \le m$ !).
Therefore we can encompass all $a_i$'s,  including their repetitions as entries of $R$,
within a two variable function $a(t,s)$ with $1 \le t,s \le n$, that is invariant under the above transformations. More precisely, for $1 \le t,s \le n$,
\[R_{v,w}=a(t,s), \quad \text{for} \quad v=(t-1)n+s \quad \text{and} \quad w=(s-1)n+t \label{R=a(ts)-1} \]
where $a(t,s)$ is a two variable complex valued function, satisfying transformations,
\[a(n-t+1,s)=a(t ,n-s+1)=a(t,s) \label{a(ts) trnasf-1} \]

Also, we can, for example, define a function $y(t,s)$, satisfying similar transformations, to encompass all $y_i$'s, including their repetitions. More precisely, for $1 \le t,s \le n$,
\[R_{v,w}=y(t,s), \quad \text{for} \quad v=(t-1)n+s \quad \text{and} \quad w=(n-s)n+t \label{R=y(ts)-1} \]
Where $y(t,s)$ is a two variable complex valued function, satisfying transformations,
\[y(n-t+1,s)=y(t,n-s+1)=y(t,s) \label{y(ts) trnasf-1} \]
 And so on for $b_i$'s, and $x_i$'s. Also for the extra $a_i$'s and $b_i$'s in the case of $n$ odd, still the same pattern holds. But,  always either $t=\frac{n+1}{2}$ or $s=\frac{n+1}{2}$ (Formulas \eqref{i j odd}-\eqref{b entries odd}). Moreover, there are no extra $x_i$'s and $y_i$'s for the case of $n$ odd. Because, for example, if  $s=\frac{n+1}{2}$, then in Formulas \eqref{R=a(ts)-1} and \eqref{R=y(ts)-1}, $v$ and $w$ coincide. Therefor, since $R_{v,w}=a(t,s)$, there is no room for  $y(t,s)$!. And so on.

We collect all the above mentioned facts in   the following lemma, for the future references.

\begin{lemma} \label{Rn2 def eqv}
The matrix $R_{n^2}$ defined in Definition \ref{Rn2 def}, could be equivalently defined as follows, for $1 \le t,s \le n$,
\[R_{v,w}=a(t,s), \quad \text{for} \quad v=(t-1)n+s \quad \text{and} \quad w=(s-1)n+t \label{R=a(ts)}\]
\[R_{v,w}=b(t,s), \quad \text{for} \quad v=(t-1)n+s \quad \text{and} \quad w=n^2+1-[(s-1)n+t] \label{R=b(ts)}\]
\[R_{v,w}=x(t,s), \quad \text{for} \quad v=(t-1)n+s \quad \text{and} \quad w=n^2+1-[(n-s)n+t] \label{R=x(ts)}\]
\[R_{v,w}=y(t,s), \quad \text{for} \quad v=(t-1)n+s \quad \text{and} \quad w=(n-s)n+t \label{R=y(ts)}\]
Where $a(t,s)$, $b(t,s)$, $x(t,s)$ and $y(t,s)$ are two variable complex valued functions, satisfying transformations,
\[a(n-t+1,s)=a(t,n-s+1)=a(t,s) \label{a(ts) trnasf}\]
\[b(n-t+1,s)=b(t,n-s+1)=b(t,s) \label{b(ts) trnasf}\]
\[x(n-t+1,s)=x(t,n-s+1)=x(t,s) \label{x(ts) trnasf}\]
\[y(n-t+1,s)=y(t,n-s+1)=y(t,s) \label{y(ts) trnasf}\]

With the following extra conditions for the case of $n$ odd. When $t=\frac{n+1}{2}$ or $s=\frac{n+1}{2}$, if $t\ne s$, we set $x(t,s)=b(t,s)$ and $y(t,s)=a(t,s)$. When $t=s=\frac{n+1}{2}$, we set
$a(t,s)=b(t,s)=x(t,s)=y(t,s)=x$.

Moreover, it is important to remember that, all other entries of $R$ which are not occupied by any $a(s,t)$, $b(s,t)$, $x(s,t)$ or $y(s,t)$, are zeros.
\end{lemma}

Now we are ready to state and to prove our main theorem.

\begin{theorem} \label{Rn2 sol yb}
For any $n$, the matrix $R_{n^2}$ defined in Definition \ref{Rn2 def} (or in Lemma \ref{Rn2 def eqv}) is a solution to Yang-Baxter equation \eqref{byb}.
\end{theorem}

In the proof of Theorem \ref{Rn2 sol yb}, we will use the following well known formulas, in which $I_{n\times n}$ is the $n$ by $n$ identity matrix, and $\mathcal{R}_{n^2\times n^2}$ could be any $n^2$ by $n^2$ matrix. These formulas describe all possible non-zero entries of $I_{n\times n} \otimes \mathcal{R}_{n^2\times n^2}$ and  $\mathcal{R}_{n^2\times n^2} \otimes I_{n\times n}$.
\[(I \otimes \mathcal{R})_{n^2(r-1)+v,\, n^2(r-1)+w}=I_{r,r}\mathcal{R}_{v,w}=\mathcal{R}_{v,w}  \label{I tens R}\]
\[(\mathcal{R} \otimes I)_{n(v-1)+r,\, n(w-1)+r}=\mathcal{R}_{v,w}I_{r,r}=\mathcal{R}_{v,w} \label{R tens I}\]

We also use the general fact that, for any three matrices, $X, Y, Z$,, if their product is defined, we have, $(XYZ)_{i,j}=\sum_{k,l}(X)_{i,k}(Y)_{k,l}(Z)_{l,j}  \label{ABC}$.

\begin{proof}{\{\textit{Theorem \ref{Rn2 sol yb}}\}}
The main idea of this proof is the following.

First, we slice $R$ (short for $R_{n^2}$) into four layers. Namely, $a$-layer denoted by $R^a$, $b$-layer denoted by $R^b$, $x$-layer denoted by $R^x$ and $y$-layer denoted by $R^y$. More precisely,
\[R=R^a +R^b +R^x +R^y \label{R layers}\]
Where, $R^a$ is the matrix containing only all $a_i$ entries of $R$ and zero elsewhere. The same goes for $R^b$, $R^x$ and $R^y$.

Next, to prove the desired equality, $(R\otimes I)(I\otimes R)(R\otimes I)=(I\otimes R)(R\otimes I)(I\otimes R)$, it is enough to prove that, for any $A$, $B$, and $C$, that could be arbitrarily chosen from the four layers of $R$,
\[(A\otimes I)(I\otimes B)(C\otimes I)=(I\otimes C)(B\otimes I)(I\otimes A) \label{R yb layers} \]

Without loss of generality, we prove this for, $A=R^a$, $B=R^x$ and $C=R^b$. Namely we will prove,
\[(R^a\otimes I)(I\otimes R^x)(R^b\otimes I)=(I\otimes R^b)(R^x\otimes I)(I\otimes R^a) \label{yb Ra Rx Rb} \]

We prove \eqref{yb Ra Rx Rb} by describing and comparing all possible non-zero ${i,j}$-entries of the left hand side ($LHS$) and the right hand side ($RHS$). Namely, we find and compare all non-zero possibilities for,
\begin{align*}
LHS&=[(I\otimes R^{a})(R^{x}\otimes I)(I\otimes R^{b})]_{i,j}=\sum_{k,l}(I\otimes R^{a})_{i,k}(R^{x}\otimes I)_{k,l}(I\otimes R^{b})_{l,j}
\end{align*}
and
\begin{align*}
RHS&=[(I\otimes R^b)(R^x\otimes I)(I\otimes R^a)]_{i,j}=\sum_{k,l}(I\otimes R^b)_{i,k}(R^x\otimes I)_{k,l}(I\otimes R^a)_{l,j}
\end{align*}

Based on Formulas \eqref{I tens R}, \eqref{R tens I}, and the definition of $R$ in Lemma \ref{Rn2 def eqv}, $(I\otimes R^{a})_{i,k}(R^{x}\otimes I)_{k,l}(I\otimes R^{b})_{l,j}$ could be non-zero, only when, by Formula \eqref{R=a(ts)},
\[i=n^2(r-1)+[(t-1)n+s] \nonumber\]
and
\[k=n^2(r-1)+[(s-1)n+t]=n[n(r-1)+(s-1)]+t \nonumber\]
which, by Formula \eqref{R=x(ts)}, enforces,
\[l=n[n^2-((n-s)n+r)]+t=n^2(s-1)+[n(n-r)+t] \nonumber\]
which in turn, by Formula \eqref{R=b(ts)}, enforces,
\[j=n^2(s-1)+[n^2+1-(n(t-1)+(n-r+1))]\nonumber\]
In such a case, by Formulas \eqref{I tens R} and \eqref{R tens I}, we will have, \newline $(I\otimes R^{a})_{i,k}(R^{x}\otimes I)_{k,l}(I\otimes R^{b})_{l,j}=a(t,s)x(r,s)b(n-r+1,s)$. Thus, we have,

\begin{align*}
LHS&=\sum_{k,l}(I\otimes R^{a})_{i,k}(R^{x}\otimes I)_{k,l}(I\otimes R^{b})_{l,j} \\
&=\sum_{t,s,r} a(t,s)x(r,s)b(n-r+1,s)
\end{align*}

With a similar method we can show that, all non-zero possibilities for the $RHS$ are,

\begin{align*}
RHS&=\sum_{k,l}(I\otimes R^b)_{i,k}(R^x\otimes I)_{k,l}(I\otimes R^a)_{l,j} \\
&=\sum_{t,s,r} a(n-t+1,s)x(n-r+1,s)b(r,s)
\end{align*}

Finally, using transformations \eqref{a(ts) trnasf}, \eqref{b(ts) trnasf} and \eqref{x(ts) trnasf}, we get $LHS=RHS$, and the proof is complete.
\end{proof}

\begin{remark}
	{\emph{It important to emphasize that, in the proof of Theorem \ref{Rn2 sol yb}, of course, there are 64 possible combined choices for all $A$, $B$, and $C$. But because of nice symmetries and similar properties among $R^a$, $R^b$, $R^x$ and $R^y$, the proof for all other choices of $A$, $B$, and $C$ will be similar to the one presented. Therefore no generality is lost in the above proof!.}}
\end{remark}

\section{Solutions to quantum Yang-Baxter equation \eqref{ayb}} \label{Solutions to algebraic Yang-Baxter equation}
Let $V$ be a $n$ dimensional complex vector (Hilbert) space. As it is well known, any solution, $R:V\otimes V \rightarrow V\otimes V$, to Yang-Baxter equation \eqref{byb} yield a solution, $\hat{R}=RS$, to the quantum Yang-Baxter equation \eqref{ayb}. Here, $S$ is the swap gate, $S:V\otimes V \to V\otimes V$, with the property that, $S(u\otimes v)=v\otimes u$, for all $u$ and $v$ in $V$.

In this section, first we give a matrix description of the swap gate $S$ for any $n$. Then we give a full description of $\hat{R}=RS$, as a solution to the quantum Yang-Baxter equation \eqref{ayb}.
It is not hard to see that the matrix, $S_{n^2}$ (or $S$ for short) defined below, is a swap gate for any dimension $n$.

\begin{definition} \label{Sn2 def}
	
For any $n$, we define the matrix, $S_{n^2}$ (or $S$ for short), by,
\[S_{v,w}=1, \,\, \text{for} \,\, v=(t-1)n+s \,\, \text{and} \,\, w=(s-1)n+t,  \,\, \text{with} \,\, 1\le t,s \le n \label{def Sn2}\]
and, $S_{v,w}=0$ elsewhere.
\end{definition}
$S_{n^2}$ for $n=2$, $n=3$ and $n=4$, are illustrated below.

{\small{
$$S_4=\left(\begin{array}{rrrr}
1 & 0 & 0 & 0 \\
0 & 0 & 1 & 0 \\
0 & 1 & 0 & 0 \\
0 & 0 & 0 & 1
\end{array}\right) \quad
S_9=\left(\begin{array}{rrrrrrrrr}
1 & 0 & 0 & 0 & 0 & 0 & 0 & 0 & 0 \\
0 & 0 & 0 & 1 & 0 & 0 & 0 & 0 & 0 \\
0 & 0 & 0 & 0 & 0 & 0 & 1 & 0 & 0 \\
0 & 1 & 0 & 0 & 0 & 0 & 0 & 0 & 0 \\
0 & 0 & 0 & 0 & 1 & 0 & 0 & 0 & 0 \\
0 & 0 & 0 & 0 & 0 & 0 & 0 & 1 & 0 \\
0 & 0 & 1 & 0 & 0 & 0 & 0 & 0 & 0 \\
0 & 0 & 0 & 0 & 0 & 1 & 0 & 0 & 0 \\
0 & 0 & 0 & 0 & 0 & 0 & 0 & 0 & 1
\end{array}\right)$$}}
{\small{
$$S_{16}= \left(\begin{array}{rrrrrrrrrrrrrrrr}
1 & 0 & 0 & 0 & 0 & 0 & 0 & 0 & 0 &
0 & 0 & 0 & 0 & 0 & 0 & 0 \\
0 & 0 & 0 & 0 & 1 & 0 & 0 & 0 & 0 &
0 & 0 & 0 & 0 & 0 & 0 & 0 \\
0 & 0 & 0 & 0 & 0 & 0 & 0 & 0 & 1 &
0 & 0 & 0 & 0 & 0 & 0 & 0 \\
0 & 0 & 0 & 0 & 0 & 0 & 0 & 0 & 0 &
0 & 0 & 0 & 1 & 0 & 0 & 0 \\
0 & 1 & 0 & 0 & 0 & 0 & 0 & 0 & 0 &
0 & 0 & 0 & 0 & 0 & 0 & 0 \\
0 & 0 & 0 & 0 & 0 & 1 & 0 & 0 & 0 &
0 & 0 & 0 & 0 & 0 & 0 & 0 \\
0 & 0 & 0 & 0 & 0 & 0 & 0 & 0 & 0 &
1 & 0 & 0 & 0 & 0 & 0 & 0 \\
0 & 0 & 0 & 0 & 0 & 0 & 0 & 0 & 0 &
0 & 0 & 0 & 0 & 1 & 0 & 0 \\
0 & 0 & 1 & 0 & 0 & 0 & 0 & 0 & 0 &
0 & 0 & 0 & 0 & 0 & 0 & 0 \\
0 & 0 & 0 & 0 & 0 & 0 & 1 & 0 & 0 &
0 & 0 & 0 & 0 & 0 & 0 & 0 \\
0 & 0 & 0 & 0 & 0 & 0 & 0 & 0 & 0 &
0 & 1 & 0 & 0 & 0 & 0 & 0 \\
0 & 0 & 0 & 0 & 0 & 0 & 0 & 0 & 0 &
0 & 0 & 0 & 0 & 0 & 1 & 0 \\
0 & 0 & 0 & 1 & 0 & 0 & 0 & 0 & 0 &
0 & 0 & 0 & 0 & 0 & 0 & 0 \\
0 & 0 & 0 & 0 & 0 & 0 & 0 & 1 & 0 &
0 & 0 & 0 & 0 & 0 & 0 & 0 \\
0 & 0 & 0 & 0 & 0 & 0 & 0 & 0 & 0 &
0 & 0 & 1 & 0 & 0 & 0 & 0 \\
0 & 0 & 0 & 0 & 0 & 0 & 0 & 0 & 0 &
0 & 0 & 0 & 0 & 0 & 0 & 1
\end{array}\right)$$ }}

Now, for any $n$, composing $R$ from Definition \eqref{Rn2 def} with $S$ from Definition \eqref{Sn2 def}, we have, $RS$, a solution to the quantum Yang-Baxter equation \eqref{ayb}.

In the following lemma we give a detailed description of $RS$ which we refer to as $\hat{R}_{n^2}$ (or $\hat{R}$ for short).

\begin{lemma} \label{Rn2 def  hat}
	For any $n$, If we set $\hat{R}_{n^2}=RS$, then the arrangement of $a_i$'s, $b_i$'s, $x_i$'s and $y_i$'s as entry positions of $\hat{R}$ is as follows,
	\[\hat{R}_{i,i}=\hat{R}_{k,k}=\hat{R}_{\tilde{k},\tilde{k}}=\hat{R}_{\tilde{i},\tilde{i}}=a_i  \label{a entries hat} \]
	\[\hat{R}_{i,\tilde{i}}=\hat{R}_{k,\tilde{k}}=\hat{R}_{\tilde{k},k}=\hat{R}_{\tilde{i},i}=b_i  \label{b entries hat} \]
	\[\hat{R}_{i,\tilde{k}}=\hat{R}_{k,\tilde{i}}=\hat{R}_{\tilde{k},i}=\hat{R}_{\tilde{i},k}=x_i  \label{x entries hat} \]
	\[\hat{R}_{i,k}=\hat{R}_{k,i}=\hat{R}_{\tilde{k},\tilde{i}}=\hat{R}_{\tilde{i},\tilde{k}}=y_i  \label{y entries hat} \]
	Where,
	\[i=(t-1)n+s \label{i hat}\]
	\[k=(t-1)n +(n-s+1) \label{k hat}\]
	\[\tilde{i}=n^2+1-i \label{i tld hat}\]
    \[\tilde{k}=n^2+1-k \label{k tld hat}\]
     For, $1 \le t,s \le m$, ($m=\frac{n}{2}$ for $n$ even and $m=\frac{n-1}{2}$ for $n$ odd). 	
	
	Plus, in the case of $n$ odd, we have the extra $a_i$'s and $b_i$'s,
		\[\hat{R}_{i,i}=\hat{R}_{j,j}=\hat{R}_{\tilde{j},\tilde{j}}=\hat{R}_{\tilde{i},\tilde{i}}=a_i  \label{a entries odd hat} \]
	\[\hat{R}_{i,\tilde{i}}=\hat{R}_{j,\tilde{j}}=\hat{R}_{\tilde{j},j}=\hat{R}_{\tilde{i},i}=b_i  \label{b entries odd hat} \]
	Where,
	\[i=(t-1)n+(\frac{n+1}{2}),\,\, j=(\frac{n-1}{2})n+t \label{i j odd hat}\]
	\[\tilde{i}=n^2+1-i,\,\, \tilde{j}=n^2+1-j\label{i j tld odd hat}\]
	 For, $1 \le t \le m$.

	Moreover, in the case of $n$ odd,  $\hat{R}_{\frac{n^2+1}{2},\frac{n^2+1}{2}}=x$.
\end{lemma}

\begin{proof}
Straightforward.
\end{proof}

\begin{remark}
{\emph{Working with $SR$ instead of $RS$, will give the same result. Because, $a_i$'s, $b_i$'s, $x_i$'s, $y_i$'s and $x$, are arbitrary and follow nice symmetries among themselves.}}
\end{remark}

Let us illustrate $\hat{R}_{n^2}$ for $n=2$, $n=3$ and $n=4$.

{\small{
$$\hat{R}_{4}=\left(\begin{array}{rrrr}
a_{1} & y_{1} & x_{1} & b_{1} \\
y_{1} & a_{1} & b_{1} & x_{1} \\
x_{1} & b_{1} & a_{1} & y_{1} \\
b_{1} & x_{1} & y_{1} & a_{1}
\end{array}\right) \quad
\hat{R}_{9}=\left(\begin{array}{rrrrrrrrr}
a_{1} & 0 & y_{1} & 0 & 0 & 0 & x_{1} & 0 & b_{1} \\
0 & a_{2} & 0 & 0 & 0 & 0 & 0 & b_{2} & 0 \\
y_{1} & 0 & a_{1} & 0 & 0 & 0 & b_{1} & 0 & x_{1} \\
0 & 0 & 0 & a_{2} & 0 & b_{2} & 0 & 0 & 0 \\
0 & 0 & 0 & 0 & x & 0 & 0 & 0 & 0 \\
0 & 0 & 0 & b_{2} & 0 & a_{2} & 0 & 0 & 0 \\
x_{1} & 0 & b_{1} & 0 & 0 & 0 & a_{1} & 0 & y_{1} \\
0 & b_{2} & 0 & 0 & 0 & 0 & 0 & a_{2} & 0 \\
b_{1} & 0 & x_{1} & 0 & 0 & 0 & y_{1} & 0 & a_{1}
\end{array}\right)$$}}

{\small{
$$\hat{R}_{16}=$$
$$\left(\begin{array}{rrrrrrrrrrrrrrrr}
a_{1} & 0 & 0 & y_{1} & 0 & 0 & 0 & 0 & 0 & 0 & 0 & 0 & x_{1} & 0 & 0 & b_{1} \\
0 & a_{2} & y_{2} & 0 & 0 & 0 & 0 & 0 & 0 & 0 & 0 & 0 & 0 & x_{2} & b_{2} & 0 \\
0 & y_{2} & a_{2} & 0 & 0 & 0 & 0 & 0 & 0 & 0 & 0 & 0 & 0 & b_{2} & x_{2} & 0 \\
y_{1} & 0 & 0 & a_{1} & 0 & 0 & 0 & 0 & 0 & 0 & 0 & 0 & b_{1} & 0 & 0 & x_{1} \\
0 & 0 & 0 & 0 & a_{5} & 0 & 0 & y_{5} & x_{5} & 0 & 0 & b_{5} & 0 & 0 & 0 & 0 \\
0 & 0 & 0 & 0 & 0 & a_{6} & y_{6} & 0 & 0 & x_{6} & b_{6} & 0 & 0 & 0 & 0 & 0 \\
0 & 0 & 0 & 0 & 0 & y_{6} & a_{6} & 0 & 0 & b_{6} & x_{6} & 0 & 0 & 0 & 0 & 0 \\
0 & 0 & 0 & 0 & y_{5} & 0 & 0 & a_{5} & b_{5} & 0 & 0 & x_{5} & 0 & 0 & 0 & 0 \\
0 & 0 & 0 & 0 & x_{5} & 0 & 0 & b_{5} & a_{5} & 0 & 0 & y_{5} & 0 & 0 & 0 & 0 \\
0 & 0 & 0 & 0 & 0 & x_{6} & b_{6} & 0 & 0 & a_{6} & y_{6} & 0 & 0 & 0 & 0 & 0 \\
0 & 0 & 0 & 0 & 0 & b_{6} & x_{6} & 0 & 0 & y_{6} & a_{6} & 0 & 0 & 0 & 0 & 0 \\
0 & 0 & 0 & 0 & b_{5} & 0 & 0 & x_{5} & y_{5} & 0 & 0 & a_{5} & 0 & 0 & 0 & 0 \\
x_{1} & 0 & 0 & b_{1} & 0 & 0 & 0 & 0 & 0 & 0 & 0 & 0 & a_{1} & 0 & 0 & y_{1} \\
0 & x_{2} & b_{2} & 0 & 0 & 0 & 0 & 0 & 0 & 0 & 0 & 0 & 0 & a_{2} & y_{2} & 0 \\
0 & b_{2} & x_{2} & 0 & 0 & 0 & 0 & 0 & 0 & 0 & 0 & 0 & 0 & y_{2} & a_{2} & 0 \\
b_{1} & 0 & 0 & x_{1} & 0 & 0 & 0 & 0 & 0 & 0 & 0 & 0 & y_{1} & 0 & 0 & a_{1}
\end{array}\right)$$ }}

\section{Unitarity conditions, Entangling property and Concluding remarks}\label{Unitarity and Entangling conditions for $R_{n^2}$}
\subsection{Unitarity and Entangling properties}
It is known that, unitary and entangling  solutions to Yang-Baxter equation, are universal quantum gates, in quantum computing \cite{brylinskis 02}. They also play a crucial role in studying the relationship between quantum entanglement and topological entanglement \cite{kaufman lomonaco 04}.

In this section, in Lemma \ref{Rn2 unitary} below whose straightforward proof is omitted, we present necessary and sufficient conditions for the matrix $R_{n^2}$ (the same for $\hat{R}_{n^2}$) to be unitary.
Next, we will, briefly, discuss the entangling property of  $R_{n^2}$. A more detailed treatment of this will appear in a sequel to this paper.

\begin{lemma} \label{Rn2 unitary}
The matrix $R_{n^2}$ (or $\hat{R}_{n^2}$) is unitary if and only if, all the following equalities are satisfied for any $i$.
\begin{align}
&a_i \bar{a_i}+b_i \bar{b_i}+x_i \bar{x_i}+y_i \bar{y_i}=1\\
&x_i \bar{a_i}+a_i \bar{x_i}+y_i \bar{b_i}+b_i \bar{y_i}=0\\
&x_i \bar{b_i}+b_i \bar{x_i}+y_i \bar{a_i}+a_i \bar{y_i}=0\\
&a_i \bar{b_i}+b_i \bar{a_i}+x_i \bar{y_i}+y_i \bar{x_i}=0
\end{align}
Plus, in the case of $n$ odd, $x \bar{x}=1$. Here, $\bar{x}$ is the complex conjugate of $x$, and the same holds for other entries.
\end{lemma}

\begin{proof}
Straightforward.
\end{proof}

As for the entangling property of $R_{n^2}$ (or $\hat{R}_{n^2}$), based on the non-entangling criteria in \cite{brylinskis 02, alagic et al 16}, one needs to show  neither $R$ nor $\hat{R}=RS$ can be factored as $X \otimes Y$ for some $X$ and $Y$. It is not hard to see this is ture, as long as $a_i$'s, $b_i$'s, $x_i$'s, $y_i$'s and $x$, not all  are zeros, and they also lack some specific polynomial relationships among them. We illustrate this for  $\hat{R}_{9}$ here.
If we wanted $\hat{R}_{9}=X \otimes Y$ for some $X$ and $Y$, it is only possible in the following way,
{\small{ $$\hat{R}_{9}=\left(\begin{array}{rrrrrrrrr}
a_{1} & 0 & y_{1} & 0 & 0 & 0 & x_{1} & 0 & b_{1} \\
0 & a_{2} & 0 & 0 & 0 & 0 & 0 & b_{2} & 0 \\
y_{1} & 0 & a_{1} & 0 & 0 & 0 & b_{1} & 0 & x_{1} \\
0 & 0 & 0 & a_{2} & 0 & b_{2} & 0 & 0 & 0 \\
0 & 0 & 0 & 0 & x & 0 & 0 & 0 & 0 \\
0 & 0 & 0 & b_{2} & 0 & a_{2} & 0 & 0 & 0 \\
x_{1} & 0 & b_{1} & 0 & 0 & 0 & a_{1} & 0 & y_{1} \\
0 & b_{2} & 0 & 0 & 0 & 0 & 0 & a_{2} & 0 \\
b_{1} & 0 & x_{1} & 0 & 0 & 0 & y_{1} & 0 & a_{1}
\end{array}\right)=$$
$\left(\begin{array}{rrr}
c & 0 & w \\
0 & d & 0 \\
t & 0 & e
\end{array}\right) \otimes
\left(\begin{array}{rrr}
f & 0 & q \\
0 & g & 0 \\
u & 0 & h
\end{array}\right)=
\left(\begin{array}{rrr|rrr|rrr}
c f & 0 & c q & 0 & 0 & 0 & f w & 0 & q w \\
0 & c g & 0 & 0 & 0 & 0 & 0 & g w & 0 \\
c u & 0 & c h & 0 & 0 & 0 & u w & 0 & h w \\
\hline
0 & 0 & 0 & d f & 0 & d q & 0 & 0 & 0 \\
0 & 0 & 0 & 0 & d g & 0 & 0 & 0 & 0 \\
0 & 0 & 0 & d u & 0 & d h & 0 & 0 & 0 \\
\hline
f t & 0 & q t & 0 & 0 & 0 & e f & 0 & e q \\
0 & g t & 0 & 0 & 0 & 0 & 0 & e g & 0 \\
t u & 0 & h t & 0 & 0 & 0 & e u & 0 & e h
\end{array}\right)$}}
From which we see, for example,  $(a_2)^2=(cg)(df)= xa_1$. Therefore if we only assume $(a_2)^2 \ne xa_1$, then $\hat{R}_{9}$ can not be factored into any product $X \otimes Y$.

We shall present a full and rigorous treatment of the entanglement property, and hence universality \cite{brylinskis 02}, of $R_{n^2}$, in a sequel to this paper.

\subsection{Discussion, Concluding remarks and Future directions}\label{Concluding remarks}

In this paper, we have created infinitely many solutions to the constant Yang-Baxter equation for any given dimension $n$. Complete solutions to Yang-Baxter equation are found only for dimension $n=2$, \cite{hietarinta, dye 03}. For dimensions $n\ge3$, the problem of complete solutions is still open. The authors in \cite{ge zhang GHZ and YB 07, rowell etal 10, rebeca chen 12}, have found very interesting solutions in other (either specific or generalized) dimensions. Our solutions are very different from theirs, not only in terms of dimensions of the matrices, but also in terms of arbitrariness of the matrix entries and their arrangements in the matrix. The authors in \cite{Shahane Khachatryan 14, Kun Hao et al 16}, using different methods, have found solutions for all $n \ge 2$ (among other things), to which our solutions feel closer relationship! However, our solutions in the current paper, are still different from theirs, in \cite{Shahane Khachatryan 14, Kun Hao et al 16}, at least for the following reasons. In our matrices, we have several more non-zero entries compared to theirs, mainly located on other diagonal (not the main diagonal) and off-diagonal. In terms of values of matrix entries, they have more distinct entries on the main diagonal, in their matrices. Also, in terms of entries' arrangements in the matrix, our of-diagonal entries are not only more than theirs in numbers, but also located differently in the matrix. One can compare ours with theirs, for example, for $n=3$.

Therefore, as far as the author's knowledge goes, the results achieved in the current paper are, novel and interesting. Not only from Mathematical point of view, but also from the perspective of Physics, and quantum computing. We hope that in future, we can generalize these results to pave the way for complete solutions to Yang-Baxter equation in any dimension $n$.
We conclude, by outlining two more future directions:

{\bf 1:} It is of great importance, and very interesting to find similar solutions to the non-constant (parameter dependent) Yang-Baxter equations. This could be done either by direct construction similar to the one used in this paper. Or by applying known {\textit{Yang-Baxterization}} methods \cite{jones baxterization 1990, ychen etal 1991}, to the results of the current paper. Thanks to Prof. Perk \cite{perk 06, perk schultz 81}, who reminded me, (after posting \cite{arash arxive 17}), of the possibility and the importance of such a leap from constant to non-constant solutions.

{\bf 2:} We believe, with a similar method, as presented in this paper, one can generate infinitely many (new) solutions to the {\textit{generalized}} Yang-Baxter equations \cite{rowell etal 10}. \newline

{{\textit{Acknowledgement:}}}
The author would like to thank Dr. Josep Batle \cite{arash-josep-2016}, for interesting and fruitful discussions during the summer of 2015, which first introduced the author to this field of research. The author also would like to thank Dr. Sinan Kapcak, for introducing him to the SageMath ({\tiny{http://www.sagemath.org/}}), which was very useful in generating large matrices and testing the validity of some formulas on them. Last but not least, I also would like to thank Mrs. Elena Ryzhova, for helping me by patiently checking large matrices, looking for some desired patterns.

\end{document}